\documentclass[preprint,times,authoryear]{elsarticle}

\usepackage{amssymb,amsmath,amsfonts,esint}

\journal{Journal of Mathematical Psychology}

\begin{document}

\begin{frontmatter}

\title{Identifiability and testability in GRT with Individual Differences}

\author[nhs]{Noah H. Silbert}
\author[rdt]{Robin D. Thomas}
\address[nhs]{Department of Communication Sciences and Disorders, University of Cincinnati}
\address[rdt]{Department of Psychology, Miami University}

\begin{abstract}
\citet{silbert_decisional_2013} showed that failures of decisional separability are not, in general, identifiable in fully parameterized $2 \times 2$ Gaussian GRT models. A recent extension of $2 \times 2$ GRT models (GRTwIND) was developed to solve this problem and a conceptually similar problem with the simultaneous identifiability of means and marginal variances in GRT models. Central to the ability of GRTwIND to solve these problems is the assumption of \textit{universal perception}, which consists of shared perceptual distributions modified by attentional and global scaling parameters \citep{soto_general_2015}. If universal perception is valid, GRTwIND solves both issues. In this paper, we show that GRTwIND with universal perception and subject-specific failures of decisional separability is mathematically, and thereby empirically, equivalent to a model with decisional separability and failure of universal perception. We then provide a formal proof of the fact that means and marginal variances are not, in general, simultaneously identifiable in $2 \times 2$ GRT models, including GRTwIND. These results can be taken to delineate precisely what the assumption of universal perception must consist of. Based on these results and related recent mathematical developments in the GRT framework, we propose that, in addition to requiring a fixed subset of parameters to determine the location and scale of any given GRT model, some subset of parameters must be set in GRT models to fix the orthogonality of the modeled perceptual dimensions, a central conceptual underpinning of the GRT framework. We conclude with a discussion of perceptual primacy and its relationship to universal perception.
\end{abstract}

\begin{keyword}
General Recognition Theory \sep identifiability \sep testability \sep GRTwIND \sep decisional separability
\end{keyword}

\end{frontmatter}

\section{Introduction}

Recent work within the general recognition theory (GRT) indicates that failures of decisional separability are not generally identifiable under common assumptions \citep{silbert_decisional_2013,thomas_technical_2014}, and it has long been known that the latent perceptual means and marginal variances are not, in general, simultaneously identifiable in GRT models \citep[e.g.,][]{wickens_maximum-likelihood_1992}. A recently developed multilevel extension of GRT (GRT with Individual Differences, or GRTwIND) has been proffered as a solution to both of these problems \citep{soto_general_2015,soto_categorization_2015}. In this theoretical note, we show that any GRTwIND model exhibiting failures of decisional separability or non-unit marginal variances is mathematically, and thereby empirically, equivalent to a model the exhibits neither trait. GRTwIND solves these two problems only \textit{conditionally}, if the assumption of of universal perception is valid, but the validity of universal perception cannot be established within GRTwIND. The purpose of this note is, in no small part, to establish precisely, and mathematically, what the assumption of universal perception entails.

It is important at the outset to mention a subtle distinction relating to the notion of identifiability. Specifically, in this paper we will distinguish between \textit{identifiability} and \textit{testability}. For the present purposes, identifiability concerns the mapping between particular parameter values and observable data given a particular model.\footnote{That is, under the assumption that the the functional form of the model and the probabilistic assumptions of the data and model parameters are true} A set of parameters is identifiable if, given a particular model, distinct parameter values map uniquely to corresponding data. On the other hand, testability concerns the relationship between the assumptions underlying a model and the model's empirical consequences. The underlying assumptions of a model are testable if relaxation of these assumptions leads to distinct empirical predictions.\footnote{Our notion of testability is closely related to the notion of \textit{structural identifiability} \citep[e.g.][]{bellman_structural_1970,eisenfeld_remarks_1985}.} Although both identifiability and testability concern the assumptions and empirical consequences of a model, the two ideas are not coextensive. We show below that there are substantial testability problems in GRTwIND, and we return to these issues periodically throughout the text as they relate to the material at hand.

We begin, in section \ref{grt}, by briefly reviewing the structure of the $2 \times 2$ Gaussian GRT model and two extensions of this model \citep[the concurrent ratings and $n \times m$ idenfication models, with $n,m>2$;][]{ashby_estimating_1988,wickens_statistical_1989,wickens_three_1992}.\footnote{Henceforth, we will use $n \times m$ to refer exclusively to \textit{identification} models with more than 2 levels on each dimension.} In section \ref{recap}, we briefly recapitulate Silbert \& Thomas's (2013) proposition $i$, which describes (a subset of) the relationships between failures of decisional separability, perceptual separability, and perceptual independence. We also recapitulate Soto et al.'s (2015) generalization of this proposition as it relates to models with multiple decision bounds on each dimension.

In section \ref{grtwind}, we describe the GRTwIND model and discuss its relationship to the concurrent ratings and $n \times m$ models. In section \ref{sub-spec}, we discuss the logic of testability with respect to Silbert \& Thomas's proposition $i$ and the assumption of universal perception in GRTwIND. We then show, through a minor generalization of proposition $i$, that any GRTwIND model with subject-specific failures of decisional separability is mathematically and empirically equivalent to a model in which decisional separability holds and in which the assumption of universal perception does not. In section \ref{unity}, we give a proof of a frequently stated, but, to the best of our knowledge, not formally proven empirical equivalence between mean and marginal variance parameters in $2 \times 2$ Gaussian GRT models. The generalization of proposition $i$ in section \ref{sub-spec} and the proof of mean-variance equivalence in section \ref{unity} together help delineate precisely what the assumption of universal perception consists of and show that this assumption is not testable with GRTwIND and associated identification-confusion data.

Finally, we discuss more general issues about the mapping between physical dimensions and modeled psychological dimensions in GRT. We propose that, in addition to the necessity of fixing the location and scale of GRT models, the \textit{dimensional orthogonality} of GRT models must be fixed, as well. We argue, following \citet{silbert_decisional_2013}, that this is typically best done by assuming decisional separability, though we also discuss other possible approaches, noting that the scope of this assumption in single-subject identification and concurrent ratings models is straightforward, whereas it is somewhat less so in multilevel models like GRTwIND and the models described by \citet{silbert_syllable_2012,silbert_perception_2014}. We conclude with a brief discussion of the relationship between universal perception and the concept of perceptual primacy.

\section{General Recognition Theory}\label{grt}

\subsection{GRT fundamentals}

GRT is a two-stage model of perception and response selection \citep{ashby_varieties_1986,kadlec_implications_1992,thomas_perceptual_2001,silbert_perception_2014,wickens_maximum-likelihood_1992}. The first stage consists of noisy perception. The second stage consists of deterministic response selection. Noisy perception is modeled with multivariate probability distributions defined over an unobserved perceptual space. Response selection is modeled with decision bounds, i.e., curves that exhaustively partition the perceptual space into response regions. Any given perceptual effect is represented as a point in perceptual space. The response to a perceptual effect is determined by the response region in which the perceptual effect occurs. The probability of a particular response to a particular stimulus is modeled as the multiple integral of the perceptual distribution corresponding to the stimulus over an appropriate response region.

\subsection{The $2 \times 2$ model}

The most common use of GRT is to analyze identification-confusion data in a $2 \times 2$ factorial paradigm, wherein the stimuli consist of the factorial combination of two levels on each of two dimensions \citep[e.g.,][]{ashby_varieties_1986,silbert_syllable_2012,silbert_perception_2014,thomas_perceptual_2001,thomas_characterizing_2001}. For example, a factorial combination of frequency and intensity could produce a set of stimulus tones that are low or high frequency and low or high intensity. In this case the four stimuli would be $A_1B_1 =$ low frequency, low intensity; $A_1B_2 =$ low frequency, high intensity; $A_2B_1 =$ high frequency, low intensity; and $A_2B_2 =$ high frequency, high intensity.

\begin{figure}[hp]
\includegraphics[width=0.75\textwidth]{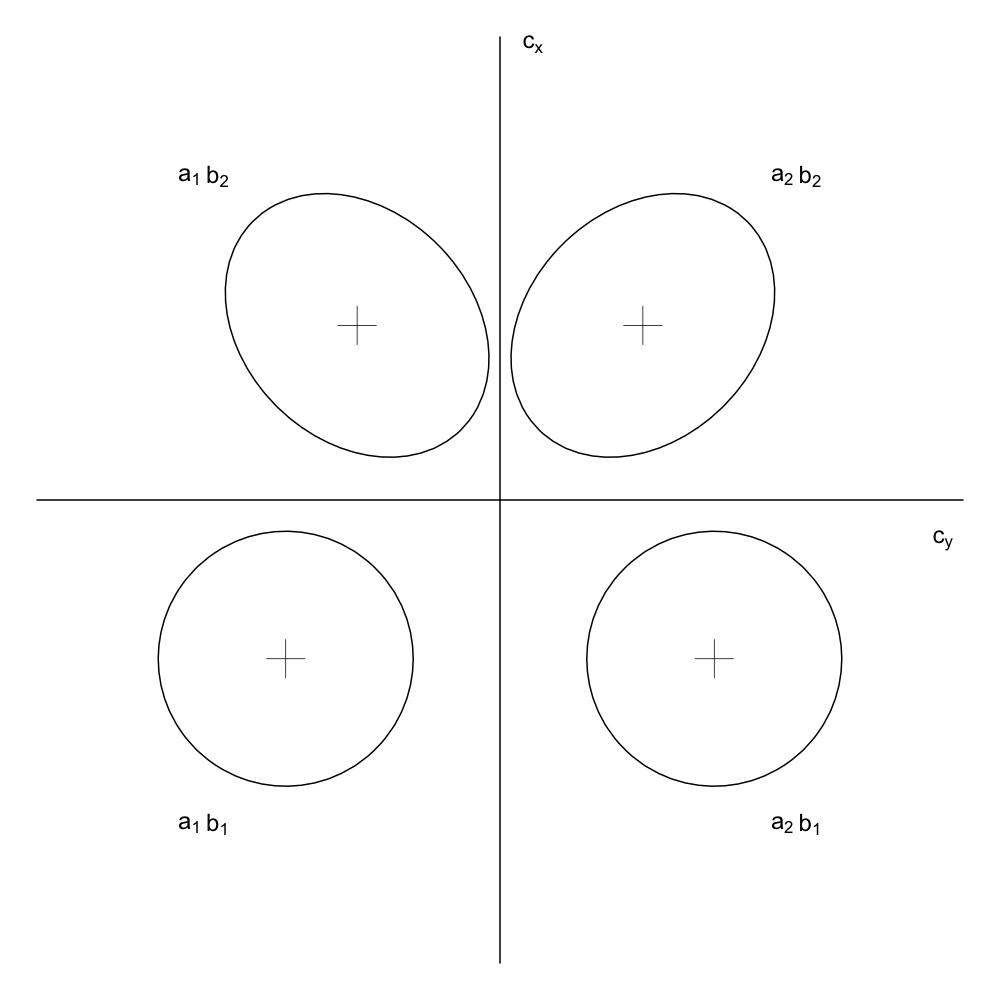}
\caption{$2 \times 2$ Gaussian GRT example illustration.}\label{example}
\end{figure}

Figure \ref{example} illustrates the equal likelihood contours and decision bounds of one possible $2 \times 2$ Gaussian GRT model. The levels of the stimuli and corresponding response regions are indicated by $a_ib_j$, with $i,j \in \{1,2\}$;\footnote{Uppercase $A_iB_j$ indicate the levels of the stimuli and corresponding perceptual distributions, while lowercase $a_ib_j$ indicate response levels.} $a_i$ indicates the level on the $x$ dimension (e.g., low vs high frequency) and $b_j$ indicates the level on the $y$ dimension (e.g., low vs high intensity). The vertical decision bound, $c_x$, partitions the $x$-axis, and the horizontal bound, $c_y$, partitions the $y$-axis. Together, they specify response regions corresponding to the same factorial structure that defines the stimuli.

This model illustrates the three dimensional interaction concepts defined in the GRT framework: perceptual independence (PI), perceptual separability (PS), and decisional separability (DS). With Gaussian perceptual distributions, PI is equivalent to zero correlation, and failure of PI is equivalent to non-zero correlation. The top two perceptual distributions illustrate failure of PI, while the bottom two exhibit PI. PS is illustrated with respect to the $y$ dimension. The perceptual distributions are perfectly horizontally aligned at each level of $B_j$; the marginal distributions of perceptual effects on the $y$ dimension do not vary as a function of the level on the $x$ dimension. By way of contrast, PS fails with respect to the $x$ dimension; the marginal perceptual distributions on this dimension vary across levels of the $y$ dimension. Finally, because the decision bounds are parallel to the coordinate axes, DS holds in this model. Decision bounds that are not parallel to the coordinate axes would represent a failure of DS.

\subsection{Multi-bound extensions of the model}

\begin{figure}[hp]
\includegraphics[width=0.75\textwidth]{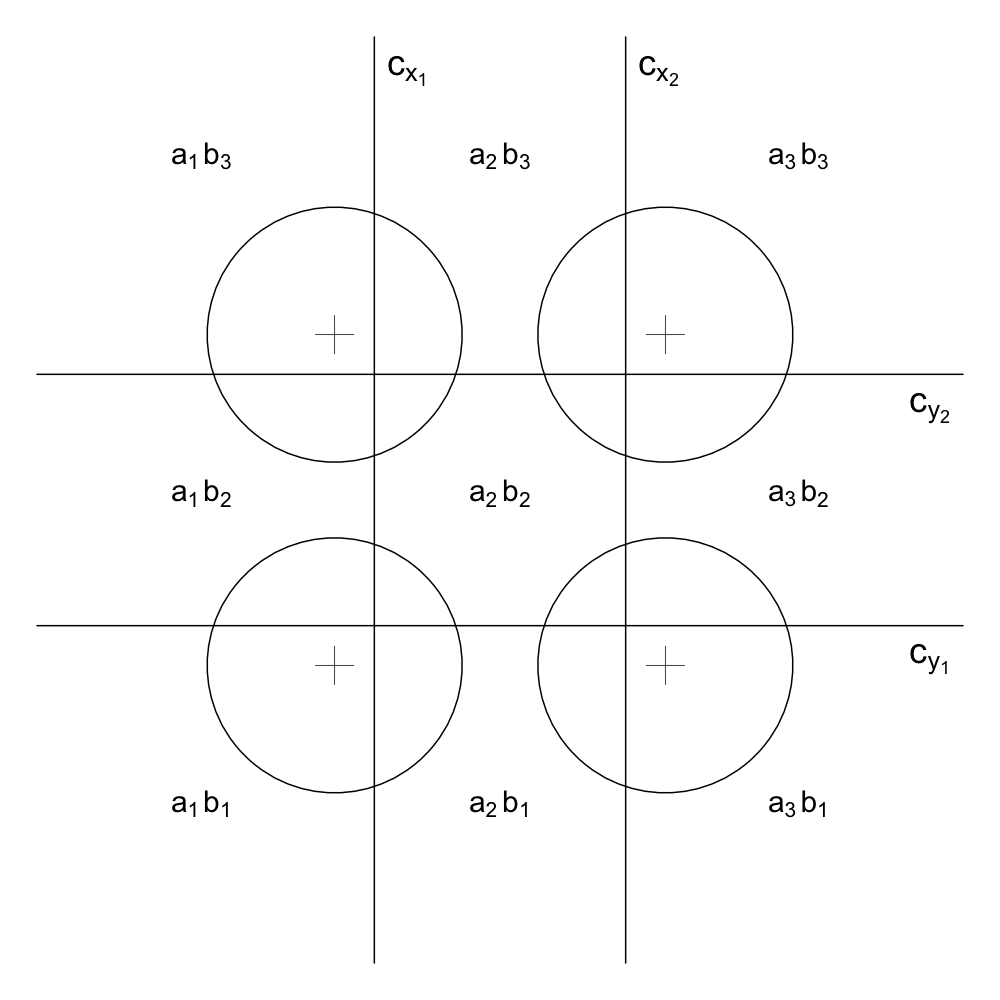}
\caption{A concurrent ratings model with two (parallel) decision bounds on each dimension.}\label{parallel}
\end{figure}

Although the $2 \times 2$ Gaussian model is the most commonly used GRT model (along with the associated factorial identification experimental paradigm), extensions of this model were developed shortly after GRT was defined as such. The concurrent ratings model \citep{ashby_estimating_1988,wickens_statistical_1989,wickens_three_1992} and the $n \times m$ model (with $n,m > 2$) were two of the first extensions of GRT, and both play an important role here. A concurrent ratings model is illustrated in Figure \ref{parallel}. The structure of this model is very similar to the $2 \times 2$ model illustrated in Figure \ref{example}, with the key difference that the concurrent ratings model has two or more decision bounds on each dimension. 

Concurrent rating models are used to analyze judgments given separately to each component (dimension) of the stimulus.\footnote{Strictly speaking, `concurrent' refers to separate responses on each dimension, while 'ratings' refers to multiple response levels on each dimension.} Crucially for our purposes, the number of response levels can be greater than those that define the stimuli. For example, experiment participants may give ratings on a $k$-point scale indicating, on each dimension, the degree to which a stimulus is judged to have been at a low or high value. The model illustrated in Figure \ref{parallel} could be used to analyze data in which subjects could respond, e.g., `low', 'uncertain', or 'high.'

We refer to a closely related extension of GRT as the $n \times m$ model. Like the $2 \times 2$ model discussed above, the $n \times m$ model is used to analyze identification data, but like the concurrent ratings model, the $n \times m$ model has multiple decision bounds on each dimension. The key difference between the concurrent ratings and $n \times m$ model is that the latter has perceptual distributions corresponding to each response region. For example, the $n \times m$ model has been used to model data from participants' identification of stimuli consisting of the factorial combination of three levels on each of two dimensions \citep[e.g.,][]{ashby_predicting_1991,thomas_multidimensional_2015}.

The $2 \times 2$, $n \times m$, and concurrent ratings models were each originally designed to analyze a single subject's data \citep{ashby_predicting_1991,thomas_perceptual_2001,thomas_characterizing_2001,wickens_statistical_1989}. Although the concurrent ratings and $n \times m$ model are distinct, for the present analyses we introduce the term \textit{multi-bound model}, which we will use to refer to both types of model in order to distinguish them as a class distinct from the standard $2 \times 2$ model. It is of central importance to the present analysis that multi-bound models predict, and the data from associated tasks may contain, responses at intermediate levels, where in the $2 \times 2$ identification identification paradigm, the data and model predictions consist only of `low' or `high' responses.

\subsection{Multilevel extensions of the model}

Two recent extensions of GRT have focused on the simultaneous analysis of multiple subjects' data. One of these extensions is a Bayesian model in which each subject's data is fit to a standard $2 \times 2$ model, while, simultaneously, the individual subjects' parameters are modeled as random variables governed by group-level parameters \citep{silbert_syllable_2012,silbert_perception_2014}. The other extension is GRTwIND, in which a group-level set of parameters are shared as well as partially modified and complemented by subject-level parameters \citep{soto_general_2015,soto_categorization_2015}. We refer to these models as \textit{multilevel} to distinguish them as a class distinct from models designed to analyze a single subject's data.

It is important to note that multi-bound and multilevel models are not mutually exclusive classes. Both the Bayesian multilevel model and GRTwIND could, in principle, be implemented as concurrent ratings or $n \times m$ models. As it happens, neither have been so implemented thus far, so, in practice, no multi-bound GRT models are multilevel, and no multilevel GRT models are multi-bound.

The distinction between multi-bound and multilevel models helps elucidate the scope of Silbert \& Thomas's proposition $i$ and Soto et al.'s generalization thereof. Specifically, \citet{silbert_decisional_2013} show that DS, PS, and PI are not simultaneously testable in single-subject $2 \times 2$ identification-confusion models. More specifically, they show that the parameters are not identifiable in a fully general Gaussian GRT model (i.e., a model in which DS, PS, and PI may all fail). \citet{soto_general_2015} generalize proposition $i$ to show that this is also true of two-dimensional Gaussian GRT models with multiple bounds on each dimension if and only if the bounds on a given dimension are parallel.\footnote{Neither \citet{silbert_decisional_2013} nor \citet{soto_general_2015} distinguish between identifiability and testability as we use the terms here, in both cases discussing these issues exclusively in terms of identifiability.} Soto et al. infer that neither proposition $i$ nor their generalization thereof apply to GRTwIND. We show below that this is incorrect.

In the following section, we briefly recapitulate, for convenience, the proof of propostion $i$.\footnote{We focus here on the non-testability of DS in $2 \times 2$ Gaussian GRT models with linear decision bounds. Although we do not address other types of decision bounds here (e.g., piecewise linear bounds), we see no reason to think that they would resolve this issue if implemented in GRTwIND. \citet{silbert_decisional_2013} show, via simulation, that the same basic issue exists in $2 \times 2$ models with piecewise linear bounds. It will be clear shortly why these results also apply to GRTwIND.} We also recapitulate Soto et al.'s generalization of this proposition.

\section{Recapitulation of \citet{silbert_decisional_2013}, proposition $i$, and Soto et al.'s generalization thereof}\label{recap}

\citet{silbert_decisional_2013} showed that single-subject $2 \times 2$ Gaussian GRT models with linear decision bounds that are not parallel to the coordinate axes can be linearly transformed to align the decision bounds with the coordinate axes. Hence, every single-subject $2 \times 2$ Gaussian GRT model with linear decision bounds exhibiting failure of DS is empirically equivalent to a single-subject $2 \times 2$ Gaussian GRT model in which DS holds. By way of contrast, models exhibiting failures of perceptual separability are not, in general, empirically equivalent to models in which perceptual separability holds.\footnote{A special case in which PS may be induced by application of linear transformations (mean-shift integrality) is given in proposition $ii$ of \citet{silbert_decisional_2013} and clarified in \citet{thomas_technical_2014}.}

Mathematically, \citet{silbert_decisional_2013} showed that a model with angle $\phi$ between $c_y$ and the $x$-axis and angle $\omega$ between $c_x$ and $c_y$ can be rotated and sheared by applying the following two linear transformations\footnote{Note that, without loss of generality, the location of the model is fixed by putting the intersection of the two decision bounds at the origin. The same rotation and shear transformations will induce DS in a non-centered model, e.g., one in which the $A_1B_1$ perceptual distribution is fixed at the origin.}:

\begin{align}
\mathbf{R} &= \left[ \begin{array}{cc}\cos\phi & -\sin\phi \\ \sin\phi & \cos\phi\end{array}\right]\\
\mathbf{S} &= \left[ \begin{array}{cc} 1 & -\frac{1}{\tan\omega} \\ 0 & 1 \end{array}\right]
\end{align}

\begin{figure}[hp]
\includegraphics[width=0.75\textwidth]{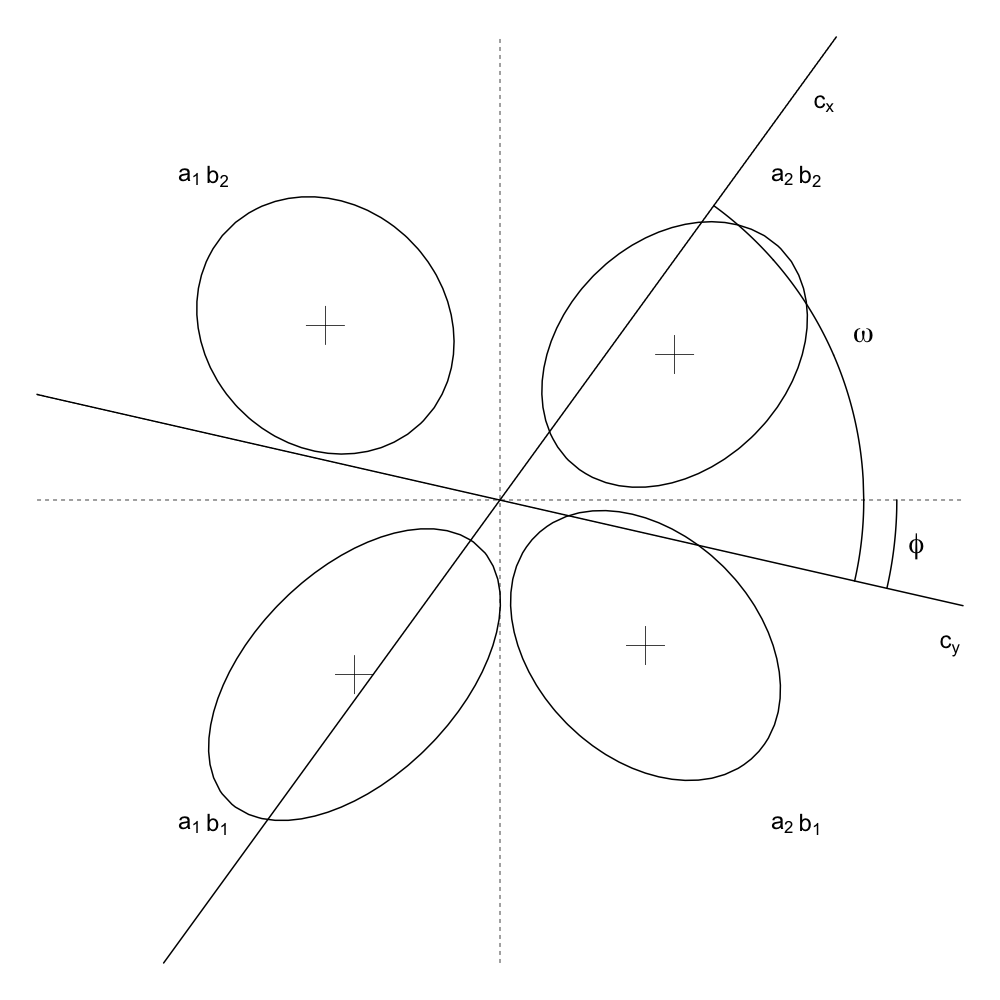}
\caption{$2 \times 2$ Gaussian GRT model exhibiting failure of DS. $\phi$ indicates the angle between the `horizontal' decision bound $c_y$ and the $x$-axis. $\omega$ indicates the angle between the two decision bounds $c_x$ and $c_y$.}\label{noDS}
\end{figure}

\begin{figure}[hp]
\includegraphics[width=0.75\textwidth]{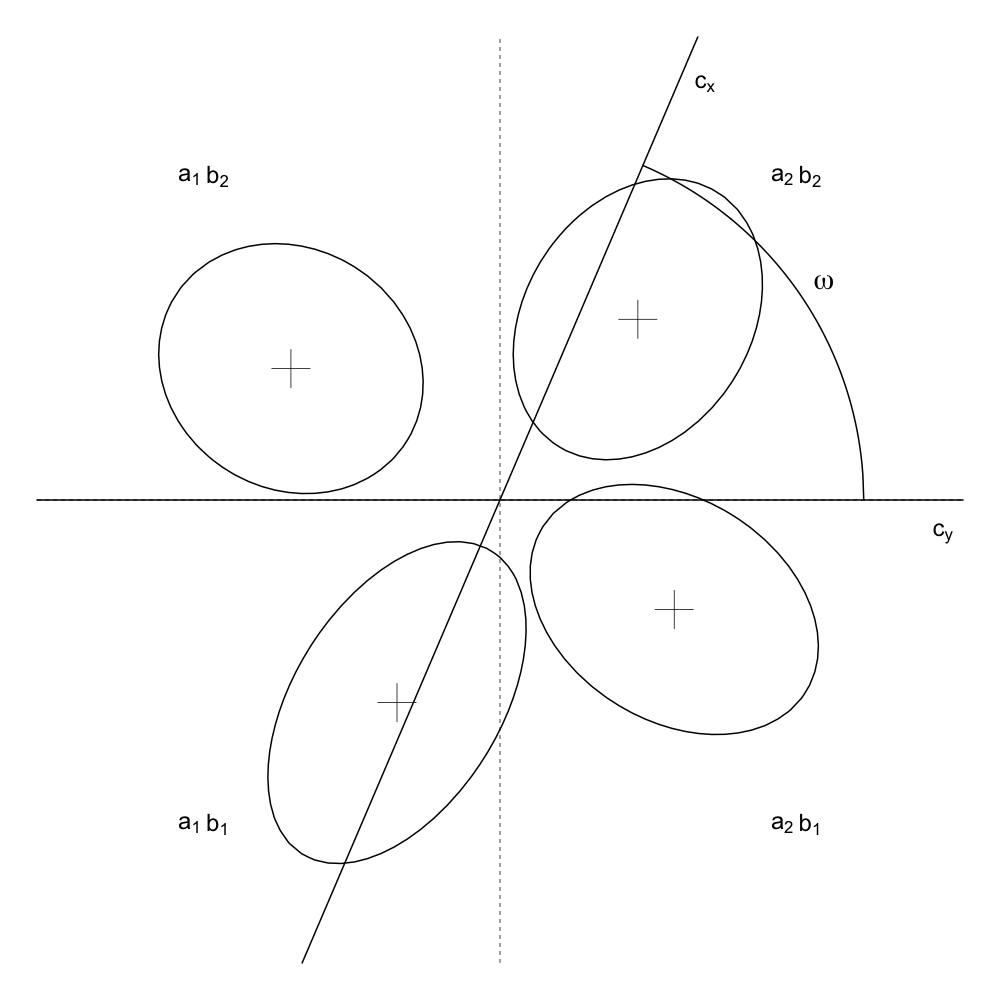}
\caption{Rotated $2 \times 2$ Gaussian GRT model.}\label{rotated} 
\end{figure}

\begin{figure}[hp]
\includegraphics[width=0.75\textwidth]{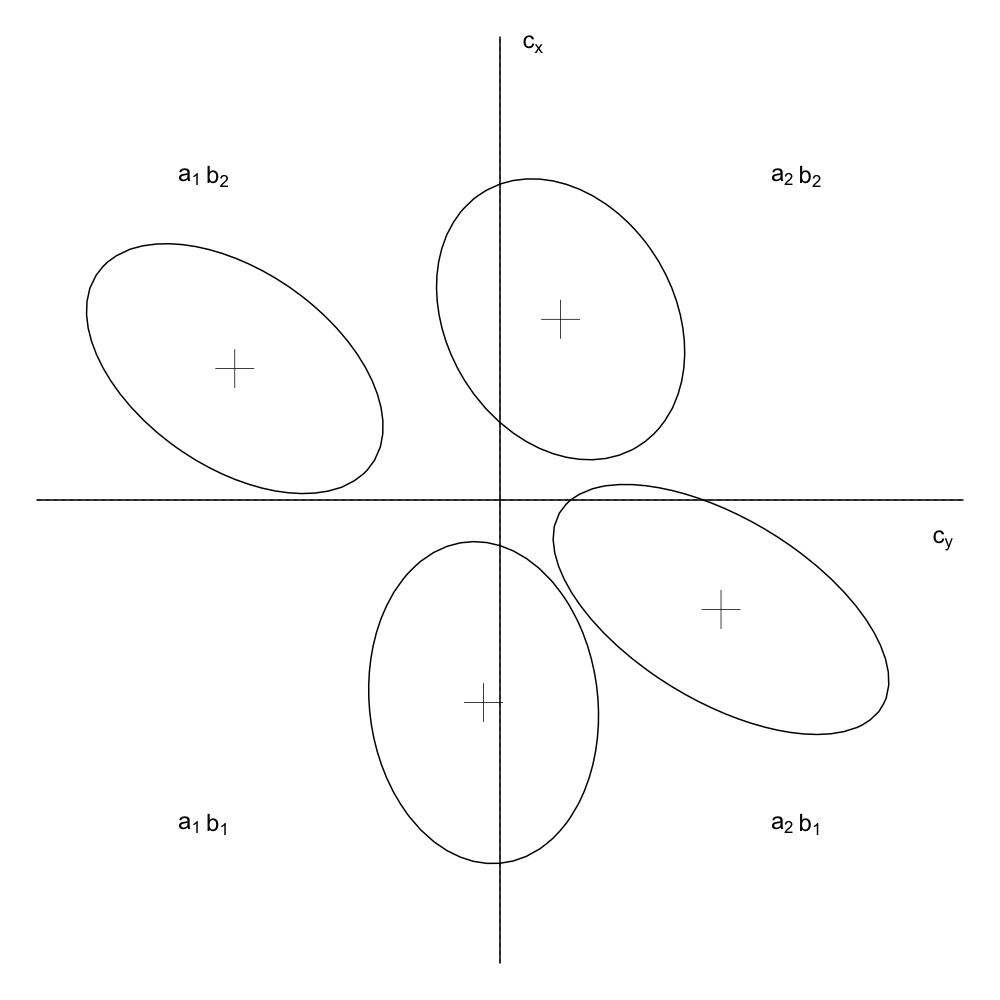}
\caption{Sheared $2 \times 2$ Gaussian GRT model.}\label{sheared}
\end{figure}

Figures \ref{noDS}, \ref{rotated}, and \ref{sheared} illustrate how a model with linear failure of DS can be rotated (via $\mathbf{R}$) and sheared (via $\mathbf{S}$) to induce DS. Figure \ref{noDS} illustrates a model exhibiting linear failure of DS, wherein the horizontal decision bound $c_y$ deviates from the $x$ axis by the angle $\phi$, and the two decision bounds are separated by the angle $\omega$.

Application of the rotation $\mathbf{R}$ aligns $c_y$ with the $x$-axis, producing the model illustrated in Figure \ref{rotated}. The angle $\omega$ between $c_x$ and $c_y$ is preserved by the rotation. Application of the shear transformation $\mathbf{S}$ preserves the alignment of $c_y$ with the $x$-axis and aligns $c_x$ with the $y$-axis, thereby inducing DS. Because these linear transformations are invertible, the predicted response probabilities are preserved \citep[][pp. 215-216]{billingsley_probability_2012}.

\citet{silbert_decisional_2013} provide a formal proof of their proposition $i$, which states that, with linear decision bounds and a single-subject $2 \times 2$ Gaussian GRT model, ``any perceptually separable but decisionally nonseparable configuration can be transformed to a configuration that is perceptually nonseparable, decisionally separable, and equivalent with respect to predicted response probabilities.'' \citet{silbert_decisional_2013} focus on the case with PS and failure of DS in order to illustrate how closely related these two notions of dimensional interaction are. However, they also note that the proposition is readily generalized to include models that do not exhibit PS prior to rotation and/or shear transformations.

\citet{soto_general_2015} generalize proposition $i$ from \citet{silbert_decisional_2013} and prove that this result also holds for Gaussian GRT models with multiple, parallel decision bounds on each dimension. More specifically, as stated above, Soto et al. show that proposition $i$ holds in two-dimensional Gaussian GRT models with more than one bound on each dimension if and only if the bounds on a given dimension are parallel.\footnote{To the best of our knowledge, it has not previously been noted that multi-bound models with non-parallel bounds on a given dimension produces uninterpretable response regions, making them incoherent models of perception and response selection. Consider, for example, the model illustrated in Figure \ref{parallel} if $c_{y_1}$ and $c_{y_2}$ were not parallel. Because non-parallel lines intersect, this would produce a region that is simultaneously \textit{above} $c_{y_2}$ and \textit{below} $c_{y_1}$.} Based on this generalization, \citet{soto_general_2015} write that Silbert \& Thomas's proposition $i$ ``is not generally true in GRT-wIND or any other model with more than one bound per dimension. The non-identifiability of decisional separability arises in such models only under very specific circumstances'' (p. 108).\footnote{Here `non-identifiability' refers to non-testability, using the terminology established above.} They conclude, incorrectly, that failures of DS are, in general, testable in GRTwIND.

Because GRTwIND is a multilevel $2 \times 2$ model, the decision bounds in GRTwIND do not function as multiple bounds on the same dimension in the same way that the decision bounds in a concurrent ratings or $n \times m$ model do; using the terminology introduced above, GRTwIND is not a multi-bound model. Soto et al. are correct that the transformations at the heart of proposition $i$ cannot induce DS simultaneously for all subjects. However, proposition $i$ applies in a subject-specific manner, such that subject-specific failures of DS in GRTwIND map one-to-one onto subject-specific rotation and shear transformations. Any GRTwIND model with universal perception and subject-specific failures of DS is mathematically and empirically equivalent to a transformed GRTwIND model with DS for all subjects and violation of universal perception.

\section{The structure of GRTwIND}\label{grtwind}

GRTwIND is a multilevel, $2 \times 2$ Gaussian GRT model that relies crucially on the assumption of universal perception \citep{soto_general_2015,soto_categorization_2015}.  Soto et al. write that ``the model assumes that the structure of the perceptual distributions is the same for all participants; that is, some aspects of perception are universal, in particular the relations between dimensions within stimuli (covariance of each distribution) and across stimuli (the means of each distribution and the ratio of their variance along a dimension).... it is also assumed that attentional and decisional processes could vary across individuals'' (p. 91).

\subsection{Group- and individual-level parameters}

Mathematically, in GRTwIND there is a shared group-level set of four bivariate Gaussian perceptual distributions. Each individual subject's modeled perceptual distributions are modifications of the shared group-level distributions. In addition, each individual subject has a set of linear decision bounds, each specified by an intercept and a slope.

The group-level perceptual distribution for stimulus $A_i,B_j$ has a mean vector and covariance matrix:
\begin{align}
\boldsymbol{\mu}_{A_iB_j} &= \left[\begin{array}{c}\mu_{x,A_iB_j} \\ \mu_{y,A_iB_j}\end{array}\right]\\
\Sigma_{A_iB_j} &= \left[\begin{array}{cc}\sigma_{xx,A_iB_j} & \sigma_{xy,A_iB_j}\\ \sigma_{yx,A_iB_j} & \sigma_{yy,A_iB_j} \end{array}\right]
\end{align}

As in the the standard $2 \times 2$ and multi-bound models, the mean vector for one distribution is set equal to $(0,0)^T$ to fix the location of the model, and, as in multi-bound models, the marginal variances in one distribution are set equal to one to fix the scale of the model.\footnote{It is typical in the standard $2 \times 2$ model to set all marginal variances equal to one. We return to this issue below.}

For individual subject $k$, the covariance matrix corresponding to stimulus $A_i,B_j$ is given by the following equation, with $\kappa_k > 0$ and $0 < \lambda_k < 1$:

\begin{equation}\label{sigma_k}
\Sigma_{k,A_iB_j} = \left[\begin{array}{cc}\frac{\sigma_{xx,A_iB_j}}{\kappa_k\lambda_k} & \frac{\sigma_{xy,A_iB_j}}{\kappa_k\sqrt{\lambda_k(1-\lambda_k)}} \\ \frac{\sigma_{yx,A_iB_j}}{\kappa_k\sqrt{\lambda_k(1-\lambda_k)}} & \frac{\sigma_{yy,A_iB_j}}{\kappa_k(1-\lambda_k)} \end{array}\right]
\end{equation}

Note that, because the (absolute and relative) scaling is applied only to marginal variances, subject $k$'s mean vector for stimulus $A_iB_j$ is just the group-level mean vector:
\begin{equation}
\boldsymbol{\mu}_{k,A_iB_j} = \boldsymbol{\mu}_{A_iB_j}
\end{equation}

Note, too, that although \citet{soto_general_2015} state, in the quote above, that the \textit{covariance} of each distribution is constant, the scaling and dimension-weighting parameters $\kappa_k$ and $\lambda_k$ ensure that this is not generally true. Rather, the assumption is that the \textit{correlation} of each distribution is constant across subjects.

More generally, it is clear that the assumption of universal perception allows for scaling of marginal variances, both with respect to the absolute scale of the space ($\kappa$) and with respect to the relative importance of each dimension ($\lambda$), but it does not allow for differences in failures of PS or PI. Naturally enough, given that it is a constraint on perceptual representations, universal perception also allows for differences with respect to failures of DS across subjects.

\subsection{GRTwIND, $n \times m$, and concurrent ratings models}

In two-dimensional, Gaussian GRT models, the predicted probability of response $a_ib_j$ to stimulus $A_kB_l$ is given by the following equation, expressed with some abuse of notation in the interest of simplicity:

\begin{equation}
\Pr(a_ib_j|A_kB_l) = \iint\limits_{R_{a_ib_j}}\mathcal{N}^{(2)}\left(\boldsymbol{\mu}_{A_kB_l},\Sigma_{A_kB_l}\right)\mathrm{d}y\mathrm{d}x
\end{equation}

Here, $\mathcal{N}^{(2)}\left(\boldsymbol{\mu},\Sigma\right)$ indicates a bivariate Gaussian (normal) probability density function\footnote{In order to maintain consistenty with \citet{silbert_decisional_2013}, we reserve $\phi$ to indicate the angle between $c_y$ and the $x$-axis. Hence, we use $\mathcal{N}$ to indicate the normal (Gaussian) probability density function.} with mean vector $\boldsymbol{\mu}$ and covariance matrix $\Sigma$, and the integration is taken over the response region $R_{a_ib_j}$.

As discussed above, in a multi-bound model for a given subject's data, a number of the response regions are determined both by decision bounds above and below (on the $y$-axis) and/or to the left and right (on the $x$ axis) of the region. See, for example, the response regions at the intermediate levels $a_2$ or $b_2$ in Figure \ref{parallel} above. Corresponding to this structure in the model, the data from a concurrent ratings or $n \times m$ identification task may contain responses at intermediate levels.

By way of contrast, no response region in the $2 \times 2$ GRTwIND model is determined by more than one decision bound on a given dimension, and the data from the corresponding $2 \times 2$ task cannot, by definition, contain responses at intermediate levels. From the perspective of subject $k$, the task is identical to the standard $2 \times 2$ factorial identification task, whether his or her data will be analyzed by GRTwIND or not.

This distinction between GRTwIND and true multi-bound models is important for understanding the scope of Silbert \& Thomas's proposition $i$ and Soto et al.'s generalization of it. It follows directly from these results that failure of DS is not generally testable in single-subject multi-bound models with parallel bounds on a given dimension. The rotation and shear transformations described by \citet{silbert_decisional_2013} apply to the whole single-subject model. But it does not then follow from \textit{this} fact that failures of DS are, in general, testable in GRTwIND. In the next section, we show that they are not.

\section{Mathematical and empirical equivalence of GRTwIND with and without decisional separability}\label{sub-spec}

\subsection{The logic of testability in GRT}

Before providing a formal demonstration of the fact that failures of DS are not, in general, testable in GRTwIND models, we discuss some of the philosophical issues underlying identifiability and testability.

As discussed above, \citet{silbert_decisional_2013} showed, in proposition $i$, that failures of DS are not testable in $2 \times 2$ GRT models. On the other hand, only a subset of failures of PS and PI are not testable. They conclude that application of $2 \times 2$ GRT models should rely on the assumption of DS. A researcher following their recommendation would be able to test failures of PS and PI, and the parameters of a Gaussian GRT model would be identifiable, \textit{conditional on the assumption that DS holds}.

Now, consider the following logic: Suppose we assume that DS holds, and we fit a $2 \times 2$ GRT model and find that PS and PI fail. Can we conclude that PS and PI have failed? It is the joint hypothesis of DS + PS + PI that has been rejected, but we do not know \textit{unconditionally} which antecedents are false. If our assumption that DS holds is not valid, then our conclusions regarding the failure of PS and PI are incorrect. The set of DS, PS, and PI together is not testable.

The logic applies in an analogous manner to GRTwIND and the assumption of universal perception. Suppose we assume that universal perception holds, then we fit a GRTwIND model and find that PS, PI, and/or DS fail. What can we conclude? In this case, it is the joint hypothesis of universal perception + PS + PI + DS that has been rejected, and, once again, we do not know unconditionally which antecedents are false. If it is universal perception, then GRTwIND provides no basis for concluding that any of the GRT interaction constructs have failed. Because of this, logically, GRTwIND does not provide a general solution to the (identifiability and testability) problems discussed by \citet{silbert_decisional_2013}. One can cover exactly the same data space with GRTwIND or with a transformed version of GRTwIND in which DS holds and failures of PS are allowed to vary across individuals.

In the following two sections, we prove that universal perception is not testable in GRTwIND by virtue of the fact that proposition $i$ implies that any GRTwIND model with subject-specific failures of DS maps one-to-one onto a model with subject-specific rotation and shear transformations in which DS holds across the board. This mathematical equivalence delineates, in part, what the assumption of universal perception consists of, and shows that a general solution to the identifiability and testability issues in question will have to be non-mathematical and not dependent on the $2 \times 2$ identification-confusion data that GRTwIND was designed to model.

\subsection{Subject-specific application of proposition $i$}\label{phi_m}

GRTwIND as a whole is, like any other GRT model, invariant to affine transformations; the modeled perceptual and decisional space is not fixed with respect to any absolute frame of reference. So, for example, rotation and/or shear transformations of a full GRTwIND model (i.e., all shared and subject-specific parameters) would preserve the full set of predicted response probabilities.

\citet{soto_general_2015} argue correctly that DS cannot, in general, be induced for all subjects simultaneously in a GRTwIND model by the application of global rotation and/or shear transformations. Their Figure 2 illustrates this fact. The argument is that, although rotation and shear transformations applied to the full model can align subject $m$'s decision bounds with the coordinate axes, as long as other subjects' decision bounds are not parallel to subject $m$'s bounds, these transformation will not also align the other subjects' bounds with the coordinate axes.

However, applying single rotation and shear transformations to the full GRTwIND model is not the only option at our disposal, nor is it a direct analog to rotation or shear transformations of single-subject $2 \times 2$ or multi-bound models. This is because \textit{each} subject $m$'s decision bound slopes define subject-specific angles $\phi_m$ and $\omega_m$ (see Figure \ref{noDS}), which define subject-specific rotation and shear matrices $\mathbf{R}_m$ and $\mathbf{S}_m$. This implies that universal perception and failure of decisional separability are not testable in GRTwIND. That is, Silbert \& Thomas's proposition $i$ applies directly to any GRTwIND model with respect to each subject's decision bounds.

The rotated and sheared mean vector for stimulus $A_iB_j$ for subject $m$ is given by:
\begin{align}\label{nu_m}
\boldsymbol{\nu}_{m,A_iB_j} &= \mathbf{S}_m\mathbf{R}_m\boldsymbol{\mu}_{A_iB_j}\\
&= \mathbf{S}_m\left[\begin{array}{c}\mu_{x,A_iB_j}\cos\phi_m-\mu_{y,A_iB_j}\sin\phi_m \\ \mu_{x,A_iB_j}\sin\phi_m+\mu_{y,A_iB_j}\cos\phi_m\end{array}\right]\nonumber\\
&= \left[\begin{array}{c}\mu_{x,A_iB_j}\cos\phi_m-\mu_{y,A_iB_j}\sin\phi_m-\mu_{x,A_iB_j}\frac{\cos\phi_m}{\tan\omega_m}+\mu_{y,A_iB_j}\frac{\sin\phi_m}{\tan\omega_m} \\ \mu_{x,A_iB_j}\sin\phi_m+\mu_{y,A_iB_j}\cos\phi_m\end{array}\right]\nonumber
\end{align}

And the rotated and sheared covariance matrix for stimulus $A_iB_j$ for subject $m$ is given by:
\begin{align}\label{sigma_m}
\Psi_{m,A_iB_j} &= \mathbf{S}_m\mathbf{R}_m\Sigma_{m,A_iB_j}\mathbf{R}^T_m\mathbf{S}^T_m\\
&= \mathbf{S}_m\Theta_{m,A_iB_j}\mathbf{S}^T_m\nonumber
\end{align}

Here, $\Sigma_{m,A_iB_j}$ is subject $m$'s scaled covariance matrix, defined in equation \ref{sigma_k} above, and $\mathbf{R}_m$ and $\mathbf{S}_m$ are subject $m$'s rotation and shear matrices, respectively.

Keeping in mind that $\left(\Theta_{m,A_iB_j}\right)_{12} = \left(\Theta_{m,A_iB_j}\right)_{21}$ (i.e., that $\Theta_{m,A_iB_j}$ is symmetric), the elements of $\Theta_{m,A_iB_j}$ are:
\begin{align}\label{Theta_m1}
\left(\Theta_{m,A_iB_j}\right)_{11} &= \frac{\sigma_{xx,A_iB_j}}{\kappa_m\lambda_m}\cos^2\phi_m-2\frac{\sigma_{xy,A_iB_j}}{\kappa_m\sqrt{\lambda_m(1-\lambda_m)}}\sin\phi_m\cos\phi_m+\frac{\sigma_{yy,A_iB_j}}{\kappa_m(1-\lambda_m)}\sin^2\phi_m\\
\left(\Theta_{m,A_iB_j}\right)_{12} &= \left(\frac{\sigma_{xx,A_iB_j}}{\kappa_m\lambda_m}-\frac{\sigma_{yy,A_iB_j}}{\kappa_m(1-\lambda_m)}\right)\cos\phi_m\sin\phi_m + \frac{\sigma_{xy,A_iB_j}}{\kappa_m\sqrt{\lambda_m(1-\lambda_m)}}\left(\cos^2\phi_m-\sin^2\phi_m\right)\\
\left(\Theta_{m,A_iB_j}\right)_{22} &= \frac{\sigma_{xx,A_iB_j}}{\kappa_m\lambda_m}\sin^2\phi_m+2\frac{\sigma_{xy,A_iB_j}}{\kappa_m\sqrt{\lambda_m(1-\lambda_m)}}\sin\phi_m\cos\phi_m+\frac{\sigma_{yy,A_iB_j}}{\kappa_m(1-\lambda_m)}\cos^2\phi_m\label{Theta_m2}
\end{align}

And keeping in mind that $\left(\Psi_{m,A_iB_j}\right)_{12} = \left(\Psi_{m,A_iB_j}\right)_{21}$ (i.e., that $\Psi_{m,A_iB_j}$ is symmetric), the elements of $\Psi_{m,A_iB_j}$ are:
\begin{align}\label{Psi_l}
\left(\Psi_{m,A_iB_j}\right)_{11} &= \left(\Theta_{m,A_iB_j}\right)_{11} - \left(\Theta_{m,A_iB_j}\right)_{12}\frac{1+\tan\omega_m}{\tan\omega_m} + \frac{\left(\Theta_{m,A_iB_j}\right)_{22}}{\tan^2\omega_m}\\
\left(\Psi_{m,A_iB_j}\right)_{12} &= \left(\Theta_{m,A_iB_j}\right)_{12} - \frac{\left(\Theta_{m,A_iB_j}\right)_{22}}{\tan\omega_m}\\
\left(\Psi_{m,A_iB_j}\right)_{22} &= \left(\Theta_{m,A_iB_j}\right)_{22}\label{Psi_m}
\end{align}

The formulas given in equations \ref{nu_m}-\ref{Psi_m} are fairly cumbersome, but they are the result of straightforward linear algebra operations. As noted above, subject $m$'s decision bound slopes are mathematically equivalent to the angles $\phi_m$ and $\omega_m$, which in turn determine $\mathbf{R}_m$ and $\mathbf{S}_m$, so the rotated and sheared model has the same number of free parameters as the specification of GRTwIND with non-zero decision bound slopes. Indeed, the rotated and sheared model is a straightforward reparameterization of the GRTwIND model, not a more general model restricted to mimic a GRTwIND model. Application of $\mathbf{R}_m$ and $\mathbf{S}_m$ to subject $m$'s parameters merely induces DS and transforms the shared mean and covariance parameters in a subject-specific manner.

These invertible, linear transformations preserve the predicted response probabilities of the model, so the GRTwIND model transformed by subject-specific rotation and shear transformations is also empirically equivalent to the original model exhibiting linear failures of DS.

\citet[][p. 93]{soto_general_2015} state that ``if violations of decisional separability are found and individual decision bounds have slightly different slopes, then it is not possible to find an equivalent model (i.e., producing the same response probabilities) in which decisional separability holds for all participants, \textit{unless the assumption of universal perception is violated} [emphasis added].'' Expressed slightly differently, subject-specific DS and shared PS and PI are only testable conditional on the validity of the assumption of universal perception. In general, the conjunction of universal perception and DS is not testable in GRTwIND, though, since relaxation of the assumption of universal perception renders the model's perceptual and decisional parameteres non-identifiable.

For every GRTwIND model, there is a mathematically and empirically equivalent model that relies on very different assumptions about the nature of the underlying perceptual and decisional interactions. We return to this issue again below.

\section{Identifiability of means and marginal variances}\label{unity}

The fact that means and marginal variances are not simultaneously identifiable in the $2 \times 2$ Gaussian GRT model was noted, in passing, more than 20 years ago \citep{wickens_maximum-likelihood_1992}. Intuitively, this makes sense as a straightforward multidimensional generalization of the constraint on the unidimensional `presence'-`absence' signal detection model, in which the variances of the noise and signal distributions are typically fixed equal to one so that the difference between the means (i.e., $d^\prime$) and a response bias parameter can both be estimated \citep{green_signal_1966}. To the best of our knowledge, however, no formal proof of this fact has been published. We provide such a proof here, after which we discuss how this result further delineates the assumption of universal perception in GRTwIND. 

\subsection{Proof of mean-variance equivalence in the standard $2 \times 2$ Gaussian GRT model}

Let $\boldsymbol{\mu}$ and $\boldsymbol{\Sigma}$ be the mean vector and covariance matrix of a bivarite Guassian density, and let $\boldsymbol{c}$ be a vector containing the response criteria\footnote{If DS holds, each decision bound is equivalent to a simple response criterion.} on each dimension:
\begin{align}
\boldsymbol{\mu} &= \left[\begin{matrix}\mu_x\\\mu_y\end{matrix}\right]\\
&\nonumber\\ 
\boldsymbol{\Sigma} &= \left[\begin{matrix}\sigma_{xx} & \sigma_{xy}\\ \sigma_{xy} & \sigma_{yy} \end{matrix} \right]\\
&\nonumber\\ 
\mathbf{c} &= \left[\begin{matrix}c_x\\c_y\end{matrix}\right]
\end{align}

If we apply the affine transformation $\mathbf{T} + \Delta$, defined below, the covariance matrix is transformed into a correlation matrix and the means are shifted with respect to the response criteria in order to preserve the distances between the means and response criteria in units of standard deviation.
\begin{equation}
\mathbf{T} + \Delta = \left[\begin{matrix}\frac{1}{\sqrt{\sigma_{xx}}} & 0\\ 0 & \frac{1}{\sqrt{\sigma_{yy}}}\end{matrix}\right] + \left[\begin{matrix}c_x -\frac{c_x}{\sqrt{\sigma_{xx}}}\\c_y - \frac{c_y}{\sqrt{\sigma_{yy}}}\end{matrix}\right]
\end{equation}

Application of this transformation produces a new covariance matrix $\mathbf{R} = \mathbf{T}\Sigma\mathbf{T}^T$ and new mean vector $\boldsymbol{\eta} = \mathbf{T}\boldsymbol{\mu} + \Delta$. The transformed covariance matrix is the correlation matrix:
\begin{align}\label{TSigT}
\mathbf{T}\Sigma\mathbf{T}^T &= \left[\begin{matrix}\frac{1}{\sqrt{\sigma_{xx}}} & 0\\ 0 & \frac{1}{\sqrt{\sigma_{yy}}}\end{matrix}\right]\left[\begin{matrix}\sigma_{xx} & \sigma_{xy}\\ \sigma_{xy} & \sigma_{yy} \end{matrix} \right]\left[\begin{matrix}\frac{1}{\sqrt{\sigma_{xx}}} & 0\\ 0 & \frac{1}{\sqrt{\sigma_{yy}}}\end{matrix}\right]\\
&=\left[\begin{matrix}\frac{\sigma_{xx}}{\sqrt{\sigma_{xx}\sigma_{xx}}} & \frac{\sigma_{xy}}{\sqrt{\sigma_{xx}\sigma_{yy}}}\\\frac{\sigma_{xy}}{\sqrt{\sigma_{xx}\sigma_{yy}}} & \frac{\sigma_{yy}}{\sqrt{\sigma_{yy}\sigma_{yy}}}\end{matrix}\right]\nonumber\\
&= \left[\begin{matrix}1 & \rho_{xy} \\ \rho_{xy} & 1\end{matrix}\right]\nonumber
\end{align}

And the transformed mean vector is the vector of response criteria added to the signed distance, in standard deviation units, between the means and response criteria:
\begin{align}\label{eta}
\boldsymbol{\eta} &= \mathbf{T}\boldsymbol{\mu} + \Delta\\ 
&= \left[\begin{matrix}\frac{1}{\sqrt{\sigma_{xx}}} & 0\\ 0 & \frac{1}{\sqrt{\sigma_{yy}}}\end{matrix}\right]\left[\begin{matrix}\mu_x\\\mu_y\end{matrix}\right] + \left[\begin{matrix}c_x -\frac{c_x}{\sqrt{\sigma_{xx}}}\\c_y - \frac{c_y}{\sqrt{\sigma_{yy}}}\end{matrix}\right]\nonumber\\
&= \left[\begin{matrix}\frac{\mu_x}{\sqrt{\sigma_{xx}}}\\\frac{\mu_y}{\sqrt{\sigma_{yy}}}\end{matrix}\right] + \left[\begin{matrix}c_x -\frac{c_x}{\sqrt{\sigma_{xx}}}\\c_y - \frac{c_y}{\sqrt{\sigma_{yy}}}\end{matrix}\right]\nonumber\\
&= \left[\begin{matrix}c_x + \frac{\mu_x-c_x}{\sqrt{\sigma_{xx}}}\\c_y + \frac{\mu_y-c_y}{\sqrt{\sigma_{yy}}}\end{matrix}\right]\nonumber
\end{align}

Before applying the transformation, the signed distance between the means and the response criteria are $\frac{\mu_x-c_x}{\sqrt{\sigma_{xx}}}$ and $\frac{\mu_y-c_y}{\sqrt{\sigma_{yy}}}$. After applying the transformation, the means are these values added to the response criteria. (The transformation applied to the response criteria produces no shift; substitute $c_x$ for $\mu_x$ and and $c_y$ for $\mu_y$ in equation \ref{eta} to see this.) Hence, the signed distances between the transformed means and the response criteria are:
\begin{align}
\eta_x - c_x &= \left(c_x + \frac{\mu_x-c_x}{\sqrt{\sigma_{xx}}}\right) - c_x = \frac{\mu_x-c_x}{\sqrt{\sigma_{xx}}}\\
&\nonumber\\
\eta_y - c_y &= \left(c_y + \frac{\mu_y-c_y}{\sqrt{\sigma_{yy}}}\right) - c_y = \frac{\mu_y-c_y}{\sqrt{\sigma_{yy}}}
\end{align}

This guarantees that the integrals of the marginal densities are equivalent pre- and post-transformation. More generally, because this transformation is invertible, it preserves the model's predicted response probabilities \citep[][pp. 215-216]{billingsley_probability_2012}. Therefore, for a pair of response criteria, there is a one-to-one mapping between empirically equivalent bivariate Gaussian GRT perceptual distributions, one of which may have arbitrary marginal variances and the other of which has unit marginal variances and suitably shifted means.

As discussed above, the non-identifiability of means and marginal variances in the $2 \times 2$ model is often addressed by setting the marginal variances of all the perceptual distributions equal to one \citep[e.g.,][]{silbert_syllable_2012,silbert_perception_2014,thomas_perceptual_2001}, and \citet{silbert_decisional_2013} fixed the marginal variances equal to one in the pre-transformation model exhibiting failures of DS.

In multi-bound models, however, the scale of the model can be established by fixing the marginal variances of just one perceptual distribution, which, along with setting the location of the model by fixing the mean vector of one distribution, allows both the means and marginal variances of the remaining distributions to be estimated. This is because the data from concurrent ratings and $n \times m$ identification tasks have more degrees of freedom than there are unknown variables (free parameters) in the corresponding models.

More specifically, suppose there are $n > 2$ and $m > 2$ rating levels in a concurrent ratings task and associated model. The data will have $4(nm-1)$ degrees of freedom,\footnote{There are $nm-1$ for each of the four stimuli, since one data value is specified if the remaining $nm-1$ are known, given the total number of times that each stimulus is presented.} while the model will have 16 parameters governing the perceptual distributions,\footnote{Three mean vectors with two free parameters each, one correlation parameter in the distribution with fixed marginal variances, three (co)variance parameters in each of the other three distributions} and $n + m - 2$ decision bound (intercept) parameters. The simplest concurrent ratings data ($n=m=3$) has 32 degrees of freedom, while the corresponding model has 20 free parameters. The degrees of freedom in the data grow multiplicatively with $n$ and $m$, while the number of free parameters in the model grows additively, so any more complex concurrent ratings data and model will have more degrees of freedom than free parameters, respectively.

Of course, a simple inequality between the degrees of freedom in the data and the number of free parameters in the model does not guarantee identifiability. We can see that such models are identifiable in this case by considering that the concurrent ratings data and model may be expressed as a system of $4(nm-1)$ equations with $14 + n + m$ unknowns of the following form:

\begin{equation}
\Pr(a_i,b_j|A_k,B_l) = \int\limits_{c_{y_{i-1}}}^{c_{y_i}}\int\limits_{c_{x_{j-1}}}^{c_{x_j}}\mathcal{N}^{(2)}\left(\boldsymbol{\mu}_{A_kB_l},\Sigma_{A_kB_l}\right)\mathrm{d}y\mathrm{d}x
\end{equation}

With $i,j,k,l \in \{1,2\}$, $c_{x_0} = c_{y_0} = -\infty$, and $c_{x_m} = c_{y_n} = \infty$. Crucially, every free parameter appears in more than one equation, since each parameter plays a role in specifying more than one predicted response probability. For example, each of the estimated decision bound partially specifies predicted probabilities on either side of the bound for every perceptual distribution, and each perceptual distribution parameter partially specifies the predicted probabilities for every response to the corresponding stimulus.

The $n \times m$ identification task and model exhibits a similar relationship, with the simplest data set having 72 degrees of freedom,\footnote{The confusion matrix is $9 \times 9$, so it has $9 \times 8$ degrees of freedom.} while the model has 57 perceptual distribution parameters\footnote{one correlation parameter in a distribution with fixed mean and marginal variances, and two mean and five (co)variance parameters in each of the other eight distributions} and $n+m-2$ decision bound parameters.\footnote{There may be additional decision bound parameters if failures of DS are modeled with piecewise linear bounds, as in, e.g., \citet{ashby_predicting_1991}, though see \citet{silbert_decisional_2013} for a discussion of some important ambiguities with the specification of piecewise failures of DS} In general, the $n \times m$ identification data and model can be expressed as a system of $nm(nm-1)$ equations with $7(nm-1)+n+m-1$ unknowns taking the same general form as the equation given for the concurrent ratings data and model above.

\subsection{GRTwIND, universal perception, and mean-variance equivalence}

In GRTwIND, there is a similar, but not identical, relationship between the data and the model. With $N$ subjects producing data in the $2 \times 2$ identification task, the data will have $12N$ degrees of freedom,\footnote{Each confusion matrix in the $2 \times 2$ task has $4 \times 3$ degrees of freedom} while the model has 16 shared perceptual distribution parameters\footnote{One correlation parameter for the distribution with fixed location and scale, and two mean and three (co)variance parameters in each of the other three distributions} and $6N$ scaling, dimension weighting, and decision bound parameters.\footnote{One scaling, one dimension weighting, two decision bound intercepts and two decision bound slopes per subject} Hence, there will be a system of $12N$ equations with $16 + 6N$ unknowns, again taking the same general form as the equation given above.

The differences between GRTwIND and multi-bound models are twofold. First, as described above, the way in which the parameters partially specify multiple predicted response probabilities differ between the two types of model. A single-subject multi-bound model is designed to analyze a single subject's data and predict intermediate (and extreme) response levels therein, whereas GRTwIND is designed to analyze multiple subjects' data and cannot, by defintion, predict intermediate response levels. Second, whereas an appropriately specified single-subject multi-bound model can be fit to a single subject's data, if the number of subjects $N \leq 2$, the number of free parameters in a GRTwIND model exceeds the degrees of freedom in the data. Hence, GRTwIND is over-parameterized with data from fewer than three subjects, as noted by \citet{soto_general_2015}.

It is also worth noting that, because GRTwIND is not a multi-bound model (i.e., because it was designed to analyze multiple subjects' $2 \times 2$ identification data), the simultaneous identification of means and marginal variances relies, like the identification of failures of DS, on the assumption of universal perception. The transformations mapping between marginal variances and means given in equations \ref{TSigT} and \ref{eta} apply in a straightforward manner to the parameters of a GRTwIND model after the application of the subject-specific rotation and shear transformation $\mathbf{R}_m$ and $\mathbf{S}_m$. Expressions for a given subject's mean, variance, and correlation parameters can be found by appropriate substitutions of terms from equations \ref{nu_m}-\ref{Psi_m} into equations \ref{TSigT} and \ref{eta}.

We can conclude from this that, in order for the assumption of universal perception to enable the simultaneous identification of means and marginal variances, it must also disallow the subject- and stimulus-specific scaling of marginal variances and means described in equations \ref{TSigT} and \ref{eta}. Given the mathematical and empirical equivalence of the covariance matrices and mean vectors on either side of equations \ref{TSigT} and \ref{eta}, it seems once again impossible that a purely mathematical justification can be found for disallowing these transformations while allowing the variance scaling described by \citet{soto_general_2015}.

\section{Conclusion}

\subsection{Testability and universal perception}

\citet{silbert_decisional_2013} showed, in their proposition $i$, that simultaneous DS, PS, and PI are not jointly testable in $2 \times 2$ Gaussian GRT models; the parameters are not identifiable in a Gaussian GRT model in which DS, PS, and PI may all fail. \citet{soto_general_2015} showed that Silbert \& Thomas's proposition $i$ holds for models with multiple decision bounds on each dimension if and only if the bounds on a given dimension are parallel. In addition, it has been known for more than 20 years that the means and marginal variances in $2 \times 2$ models are not both identifiable, though they are identifiable in concurrent ratings and $n \times m$ identification models, which we here refer to as multi-bound models \citep{ashby_predicting_1991,ashby_estimating_1988,wickens_maximum-likelihood_1992}. A recent multilevel extension of GRT called GRTwIND was developed in an attempt to solve these problems in the $2 \times 2$ case \citep{soto_general_2015,soto_categorization_2015}.

Soto and colleagues argue that if the assumption of universal perception is valid, then GRTwIND solves both problems. As described by \citet{soto_general_2015}, universal perception constrains the GRTwIND model so that the nature of any perceptual interactions is common to all subjects. In practice, this means that the model has a single set of perceptual distributions parameterized by mean vectors and covariance matrices. The full GRTwIND model adds to these shared parameters a set of subject-specific scaling parameters $\lambda_m$ and $\kappa_m$, $m \in \{1,2,\dots,N\}$, which modify the perceptual covariance matrices, and subject-specific decision bounds, each of which is specified by intercept and slope parameters.

In section \ref{phi_m}, we showed that each subject's decision bound slopes map one-to-one onto subject-specific angles $\phi_m$ and $\omega_m$, which in turn define subject-specific rotation and shear matrices $\mathbf{R}_m$ and $\mathbf{S}_m$ (see equations \ref{nu_m}-\ref{Psi_m}). These one-to-one mappings prove that GRTwIND with subject-specific failures of decisional separability is mathematically, and thereby empirically, equivalent to a model in which decisional separability holds for all subjects and in which universal perception is violated. Finally, we showed that means and marginal variances are not, in general, simultaneously identifiable in $2 \times 2$ GRT models, including the GRTwIND model transformed by subject-specific rotations and shears.

Universal perception is defined as shared (failures of) perceptual independence and perceptual separability, but none of the dimensional interactions defined in the GRT framework are directly observable. Indeed, the greatest strength of GRT is its utility in allowing us to draw inferences about unobservable dimensional interactions from observable data. The mathematical facts described above delineate precisely what universal perception must consist of. Per the original description of GRTwIND, universal perception allows subject-specific marginal variance scaling. The results described in this paper indicate that universal perception must also \textit{disallow} the subject-specific rotation and shear transformations described in equations \ref{nu_m}-\ref{Psi_m} and the subject- and stimulus-specific mean and marginal variance scaling transformations described in equations \ref{TSigT} and \ref{eta}.

These results establish the complete mathematical and empirical equivalence of GRTwIND and a model with subject-specific rotation and shear transformations. Hence, the pattern of allowed and disallowed transformations described above can only be justified by non-mathematical means or by empirical means other than the $2 \times 2$ identification data that GRTwIND was designed to model. Any possible validation of the assumption of universal perception depends on such justification. Of course, validation of universal perception may one day be found in data from other tasks and models.

\subsection{Dimensional orthogonality and perceptual primacy}

We conclude by proposing that the full suite of results concerning the (lack of) identifiability and testability of DS, PS, and PI in GRT models, and of universal perception in GRTwIND, points toward an important, and thus far incompletely addressed issue at the heart of the GRT framework, namely the orthogonality of the modeled perceptual dimensions. From the initial development of GRT, it was recognized that orthogonality of perceptual dimensions is intimately intertwined with perceptual and decisional dimensional interactions \citep{ashby_varieties_1986}. Indeed, \citet{ashby_varieties_1986} discuss the difficulties related to testing dimensional orthogonality in some detail. Nonetheless, the full import of this assumption seems only now, three decades later, to be fully understood.

As discussed above, in order for a GRT model's parameters to be identifiable, the location and scale of the model must be fixed, and this is typically done by setting one mean vector equal to the origin and by setting one perceptual distribution's marginal variances equal to one. The recent mathematical developments in the GRT framework, including those discussed above, indicate that we must also fix the orthogonality of the perceptual dimensions.

Without describing it explicitly in these terms, \citet{silbert_decisional_2013} recommend fixing the orthogonality of the perceptual dimensions by assuming that decisional separability holds in a single-subject $2 \times 2$ model. We assume that it would also be possible to fix dimensional orthogonality by constraining a subset of perceptual distribution parameters (e.g., by setting $\mu_{x,A_1B_1} = \mu_{x,A_1B_2}$ and $\mu_{y,A_1B_1} = \mu_{y,A_2B_1}$). However, as noted by \citet{silbert_decisional_2013}, decisional separability can always be induced in the $2 \times 2$ model, whereas perceptual separability can only be induced via linear transformations from a narrowly circumscribed subset of failures of perceptual separability. Any constraints on perceptual distribution parameters serving to fix dimensional orthogonality should be carefully designed to take these facts into account. Assuming that decisional separability holds has the benefit of being simple to implement and understand, though we acknowledge that arguments based on simplicity do not provide an overwhelmingly strong rationale for preferring one or another approach to fixing dimensional orthogonality.

The analysis described above can be interpreted as another reflection of the need to fix the orthogonality of the perceptual dimensions in GRT models. As with the standard $2 \times 2$ and single-subject multi-bound models, the location, scale, and orthogonality must be fixed in GRTwIND. Also as with the standard $2 \times 2$ and multi-bound models, it seems simplest to us to ensure orthogonality by fixing decision bounds to induce decisional separability. Although it may be possible to find a suitable restriction on a subset of perceptual distribution parameters to fix dimensional orthogonality in GRTwIND, the fact that perceptual parameters are shared across subjects seems likely to complicate matters. Again, though, it is important to keep in mind that neither simplicity nor interpretability provide anything more than a pragmatic justification for fixing orthogonality by constraining decisional rather than perceptual parameters. 

It's worth noting that the other recent multilevel extension of GRT is affected by the need to fix the orthogonality of the dimensions even more strongly than is GRTwIND. \citet{silbert_syllable_2012,silbert_perception_2014} used a multilevel model in which each subject's data is modeled by a fully parameterized $2 \times 2$ Gaussian GRT model, with group level parameters governing variation across subjects with respect to each subject-level parameter. Whereas GRTwIND may solve two important GRT-specific testability and identifiability problems if the assumption of universal perception holds, the multilevel model described by \citet{silbert_syllable_2012,silbert_perception_2014} cannot solve either, regardless of the validity of any underlying assumptions.

With respect to universal perception, our results showing that GRTwIND is mathematically equivalent to a model in which decisional separability holds for all subjects (section \ref{sub-spec}) and in which all marginal variances are equal to one (section \ref{unity}) go beyond the issue of dimensional orthogonality. Specifically, the assumption of universal perception, which consists of a strong set of constraints on allowable subject-specific modifications of perceptual distribution properties, seems to be concerned less with dimensional \textit{orthogonality} and more with dimensional \textit{primacy}.

Establishing the perceptual primacy of a particular set of dimensions demands evidence that is not simple to come by. For example, \citet{melara_perceptual_1990} argue that patterns of change in the magnitude of Garner interference across levels of physical dimension orientations provide evidence of perceptual primacy (or lack thereof), but they analyzed the perception of well-defined (orthogonal) physical dimensions (acoustic frequency and intensity). By way of contrast, assuming perceptual primacy, \citet{soto_general_2015} analyzed  dimensions with no straightforward physical definitions (facial identity and neutral vs sad emotional expressions), and \citet{soto_categorization_2015} analyzed novel dimensions based on morphed faces, stating that ``there are no psychologically-meaningful directions in a space constructed this way'' (p. 110). Similarly, \citet{silbert_syllable_2012,silbert_perception_2014} used GRT to probe interactions between dimensions defined with respect to abstract linguistic categories.

Evidence for perceptual primacy with respect to novel dimensions may be particularly difficult to find, as unsupervised learning seems to play a role in the creation of ad-hoc perceptual dimensions \citep{jones_structure_2013}. When considering the primacy of particular dimensions and assumptions of universal perception, it is also worth keeping in mind that holistic vs analytic cognition may vary across cultures \citep{nisbett_culture_2001}.

To the extent that dimensional primacy and/or universal perception requires shared perceptual correlations across subjects, one could argue against the rotation and shear transformations described above. However, it is not clear that the shared perceptual correlations of Soto et al.'s universal perception is a valid assumption in all cases. For example, multilevel $2 \times 2$ GRT models fit to speech perception data exhibit substantial variation of perceptual distribution correlations across subjects \citep{silbert_syllable_2012,silbert_perception_2014}. Similarly large differences in correlations across subjects have been reported in $2 \times 2$ and $3 \times 3$ identification data from face recognition tasks \citep{thomas_perceptual_2001,thomas_multidimensional_2015}. The assumption that the means and variance ratios are constant across subjects seems to be similarly suspect \citep[e.g.,][]{silbert_syllable_2012,silbert_perception_2014,thomas_multidimensional_2015}.

In conclusion, it seems clear to us that an independent validation of the assumption of universal perception as originally described by \citet{soto_general_2015}, and as elaborated on here, would represent important progress in the GRT framework.

\section{References}
\bibliographystyle{elsarticle-harv}
%\bibliography{silbert_library_zotero_25July2016}

\end{document}